\documentstyle[12pt]{article}
\newcommand{\etal}{{\it et al. }}

\newcommand{\ads}{accretion disks }
\newcommand{\beq}{\begin{equation} }
\newcommand{\eeq}{\end{equation} } 
\newcommand{\bcn}{\begin{center} }
\newcommand{\ecn}{\end{center} }
 
\begin{document}
\begin{center}
%\begin{Large}
{\bf GENERALIZED THERMAL INSTABILITY CRITERION OF BLACK HOLE ACCRETION DISKS}\footnote{Supported 
by the Postdoctoral Science Foundation of China}\\[3mm]
WU Xue-bing $^{1,2}$\\
1. Beijing Astronomical Observatory, Chinese  Academy of Sciences, Beijing 100080, China\\
2. Institute of Theoretical Physics, Chinese  Academy of Sciences, Beijing 100080, China\\[3mm]
%\end{Large}
%\end{center}
\bigskip
%{\small{\bf ABSTRACT}}
\end{center}
%\begin{small}
\begin{quotation}
{\it The conventional thermal instability criterion can not be applied to the 
 advection-dominated accretion disks around black holes where the radiative cooling is 
insufficient to balance the viscous heating. The surface density change
associated with the temperature perturbations, which was usually neglected in
deriving the conventional criterion, was recently shown to be much significant
in the advection-dominated disks. Considering both advection and surface
density change, I suggested a generalized thermal instability criterion. 
By applying it to the optically thin and optically thick 
advection-dominated disks, I found that the former one is thermally stable and the
latter one is thermally unstable against short wavelength perturbations, which agrees 
well to those found
	recently by both analytic and
quantitative stability analyses.}\\[3mm]
%\end{small}
%\keywords{accretion, accretion disks; black hole physics; instabilities}
\end{quotation}
%\end{abstract}
%
%{\large {\bf 6.1 Introduction}}

Accretion disk around black holes can provide a large amounts of released
energy and is thus believed to be the source of high energy emission of Galactic black hole
X-ray binaries and active galactic nuclei $^1$.
The stability properties of accretion disks are very important because a
global violently unstable disk may not exist in nature and some instabilities
restricted in a certain region of the disk may contribute to the observed light
variations in many active astronomical systems. Shortly after the construction 
of the standard model of an optically thick, 
geometrically thin accretion disk $^2$, the inner region 
of such a disk was found to be both thermally and secularly unstable $^{3,4}$.
  The hot optically thin disk model,
proposed by Shapiro, Lightman \& Eardley $^5$ to explain the hard X-ray 
spectra of black hole X-ray binary Cyg X-1, was also proved to be thermally unstable $^6$.
The general stability criteria  of \ads were derived by
Piran $^7$ who considered various kinds of viscosity laws and cooling processes.
 These criteria can be used as an {\it a priori}
check if some choice of cooling mechanisms gives rise to a stable disk even before
the complete disk structure equations are solved. The thermal instability
criterion can be expressed as $^8$:
%\bcn
\beq
(\frac{\partial lnQ^-}{\partial lnT})_{\Sigma}<(\frac{\partial lnQ^+}{\partial 
lnT})_{\Sigma},
\eeq
%\ecn
where $Q^+$, $Q^-$ and T are the viscous heating rate, radiative cooling rate
and temperature. $\Sigma$ is the surface density defined as $\Sigma=\rho H$ 
where $\rho$ and $H$ are the density and disk height respectively. Such a
criterion is identical to that appeared in Piran $^7$ if we relace $T$ with 
$H$. The thermal instability can be well understood according to this criterion.
Considering a small temperature increase and assuming that the surface density 
does not change 
within the thermal time scale, if the viscous heating rate increases more rapidly 
than the radiative cooling rate does (as indicated in inequality (1)), it will lead 
to a further increase of 
temperature and the thermal instability sets in. If the standard viscosity
law is adopted, Piran has pointed out that
the inner region of accretion disks is always thermally unstable no
matter what kind of cooling mechanism is involved $^7$ .

However, the thermal instability criterion we mentioned above can not be applied
directly to the advection-dominated disks. In the advection-dominated disk models,
the radiative cooling is inefficient and most of the  energy generated by viscous heating
is transfered to increase the entropy of accretion flow and then is advected inwards onto 
the black hole $^{9,10,11}$. To derive a criterion for the thermal 
instability of 
advection-dominated disks, we should consider the advected energy in the
energy conversation equation. Moreover, the surface density change associated
with the temperature perturbations, which used to be negligible in previous
stability analyses of the geometrically thin disks, should be considered in deriving
 a thermal instability
criterion of advection-dominated disks. Some recent works on the stability of 
advection-dominated disks
have indicated that the surface density change is much significant in
these disks comparing to the minor pressure change~$^{12,13,14}$. In this letter,
I will present a generalized thermal instability criterion by considering both
advection ad surface density change.

In deriving the conventional thermal instability criterion, the energy balance
between heating and cooling was assumed in the unperturbed case. However, in an
advection-dominated disk, this balance is realized among viscous heating,
radiative cooling and advection (for simplicity, the effects of
magnetic field and thermal diffusion are not included in the present study). That is,
%\bcn
\beq
Q^+ = Q^- + Q^{ad}
\eeq
%\ecn
where $Q^+$, $Q^-$ and $Q^{ad}$ are the viscous heating rate, radiative cooling rate
and the energy advection rate respectively. The crucial point of the thermal 
stability of accretion disks is to examine whether or
not this balance is still held if the temperature perturbations are involved.
To do this examination, we first write some relative energy changes as follows:
\beq
%\begin{array}
\frac{\delta Q^+}{Q^+}=x_1\frac{\delta T}{T}+y_1\frac{\delta\Sigma}{\Sigma},~~~
\frac{\delta Q^-}{Q^-}=x_2\frac{\delta T}{T}+y_2\frac{\delta\Sigma}{\Sigma},~~~
\frac{\delta Q^{ad}}{Q^{ad}}=x_3\frac{\delta T}{T}+y_3\frac{\delta\Sigma}{\Sigma}
%\end{array}
\eeq
where the relative changes of $Q^+$, $Q^-$ and $Q^{ad}$ are expressed as
the functions of temperature  and  surface
density. $x_i$ and $y_i$ ($i=1,2,3$) can be defined as:
%\begin{array}
$$
x_1=(\frac{\partial lnQ^+}{\partial lnT})_{\Sigma},~~y_1=(\frac{\partial lnQ^+}
{\partial ln\Sigma})_T;~~~
x_2=(\frac{\partial lnQ^-}{\partial lnT})_{\Sigma},~~y_2=(\frac{\partial lnQ^-}
{\partial ln\Sigma})_T;$$
\beq
x_3=(\frac{\partial lnQ^{ad}}{\partial lnT})_{\Sigma},~~y_3=(\frac{\partial 
lnQ^{ad}}{\partial ln\Sigma})_T
%\end{array}
\eeq
If we also define $q=Q^{ad}/Q^+$ and $Q^-=(1-q)Q^+$, we can obtain:
\beq
\frac{\delta Q^+ -\delta Q^- -\delta Q^{ad}}{Q^+}=[(x_1-x_3)-(1-q)(x_2-x_3)]
\frac{\delta T}{T}+[(y_1-y_3)-(1-q)(y_2-y_3)]\frac{\delta \Sigma}{\Sigma}
\eeq
The thermal instability will arise if the left side of above equation
is positive, which means that with a small increase of temperature the viscous 
heating rate grows more fast than the
sum of radiative cooling rate and energy advection rate does. Therefore, a thermal
instability criterion including the advection term and the surface density 
change will be:
\beq
x_1-(1-q)x_2-qx_3>-\delta_{\Sigma T}[y_1-(1-q)y_2-qy_3]
\eeq
where $\delta_{\Sigma T}\equiv \frac{\delta\Sigma}{\Sigma}/\frac{\delta T}{T}$, which
describes the relative changes of surface density to the temperature 
perturbations. This criterion can be regarded as the generalization of the
conventional criterion (inequality (1)). If we keep the surface density as
a constant and take $q$ as zero, it will be the identical to the inequality (1). Because the
advection term and the surface density change  are considered in our new 
criterion, we can use it now to check the thermal instability of 
advection-dominated disks.

For an optically thin disk, the pressure is
dominated by gas and the opacity is dominated by electron scattering $^{10}$. 
We assume that radiative cooling is provided
by
thermal bremsstrahlung with emissivity as
\beq 
Q^-=1.24\times10^{21}H\rho^2T^{1/2}ergss^{-1}cm^{-2}.
\eeq
The viscous heating rate is given by the standard formula 
\beq
Q^+=\Sigma\nu(r\frac{\partial\Omega}{\partial r})^2,
\eeq
Where $\nu$ is the viscosity parameter and $\Omega$ is the angular velocity. The advection cooling rate is taken as
\beq
Q^{ad}=C_v[\Sigma V_r\frac{\partial T}{\partial r}-(\Gamma_3-1)
TV_r\frac{\partial \Sigma}{\partial r}],
\eeq
 where
 $C_v=3c_s^2/2T$, $\Gamma_3=5/3$, $V_r$ is the radial velocity and $c_s$ is the local sound speed. 
Considering $p=K\rho T$,
$H=c_s/\Omega_k$, $c_s^2=p/\rho$, $\nu=\alpha c_sH$ and
$V_r\sim \nu/r$, we have $Q^+\propto\Sigma T$, $Q^-\propto
\Sigma^2$ and $Q^{ad}\propto\Sigma T^2[s(T)-s(\Sigma)]$. Here
$s(T)\equiv\frac{d lnT}{d lnr}$ and
$s(\Sigma)\equiv\frac{d ln \Sigma}{d lnr}$ and
we assume them to be constant, which is valid if $T$ and
$\Sigma$ are the power law functions of $r$ as indicated, for example,
by the self similar solutions $^{9}$.
Therefore, for an optically thin disk we have
$x_1=1$, $y_1=1$, $x_2=0$, $y_2=2$, $x_3=2$ and $y_3=1$.
Taking these values into inequality (6), we can obtain a
criterion for thermal instability of an optically thin disk, that is
\beq
\delta_{\Sigma T}<1-\frac{q}{1-q}.
\eeq

If the optically thin disk is radiative cooling dominated, which means
$q\sim 0$, above inequality requires $\delta_{\Sigma T}<1$. In the
long perturbation wavelength case, there is no appreciable surface
density change so we have $\delta_{\Sigma T}\sim 0$. In
the short perturbation wavelength case, a recent study by Wu $^{14}$
showed  $-1<\delta_{\Sigma T}<0$. Therefore, our criterion proves
again that an optically thin disk is thermally unstable if it is
radiative
cooling dominated. On the other hand, if the optically thin disk is advection-dominated,
which means $q\sim 1$, inequality (10) requires $\delta_{\Sigma
T}<-\infty$.
Since we always have $-1<\delta_{\Sigma T}<0$ for an optically thin, advection-dominated
disk,
obviously this inequality means that the disk is thermally stable.

If the optically thin disk is two-temperature one, the radiative
cooling
may be dominated by Comptonization through the loss of energy of
electrons. Then the radiative cooling takes the form of $^5$:
\beq
Q_- =(4kT_e /m_e c^2)\rho H \kappa_{es} U_r c.
\eeq
where $m_e$, $T_e$ are the mass and temperature of electron, $\kappa_{es}=0.4cm^2g^{-1}$ 
is the electron 
scattering opacity. $U_r$ is the radiation energy density of soft photons, which
we assume, for simplicity, does not change on time scale shorter compared to 
$\Omega^{-1}$. Because the ions and electrons are coupled by collisional 
energy exchange, the 
loss of energy of electrons can be balanced by the energy capture from ions. 
The 
exchange rate is
\begin{equation}
Q_{i-}=(3/2)\rho H \nu_E k(T_i-T_e)/m_p
\end{equation}
where $\nu_E$ is the electron-ion coupling rate, and can be approximated by 
$\nu_E=2.4\times 10^{21}ln \Lambda {\rho}{T_e}^{-3/2}$  where the Coulomb 
logarithm $ln \Lambda$ is about 15. The ion temperature $T_i$
is usually one or two orders higher than the electron temperature $T_e$. 
Taking $Q_-=Q_{i-}$ and $T_i >> T_e$, we can 
get $Q_-\propto \Sigma^{7/5}{T_i}^{1/5}$. Then we have $x_2=1/5$ and
$y_2=7/5$, then the thermal instability criterion now becomes:
\beq
\delta_{\Sigma T}<2-\frac{5q}{2(1-q)}.
\eeq
Similar as we discussed above, this inequality clearly shows that a radiative 
cooling dominated two-temperature
disk
is thermally unstable and an advection-dominated two-temperature disk
is
thermally stable. A more recent quantitative stability analysis also
reaches
the same conclusion $^{15}$. Moreover, from inequalities (10) and
(13), we note that in the long wavelength perturbation case where
$\delta_{\Sigma T} \sim 0$,  the bremsstrahlung cooling disk and the two temperature
disk may become thermally stable when $q > \frac{1}{2}$ and $q > \frac{4}{9}$, 
respectively.

For an optically thick accretion disk, if it is dominated by radiation pressure and
electron scattering, we can write the viscous heating rate as
$ Q^+=\Sigma\nu(r\frac{\partial\Omega}{\partial r})^2 \propto 
\frac{T^8}{\Sigma}$ and the radiative cooling rate as 
$Q^-=\frac{4acT^4}{3\kappa \Sigma}\propto \frac{T^4}{\Sigma}$ where
$\kappa$ is the opacity dominated by electron scattering. The
advection
cooling rate, which is described by equation (9), can be expressed as
$Q^{ad}\propto \frac{T^{16}}{\Sigma^{3}}[s(T)-s(\Sigma)]$ for an
optically
thick disk. Therefore, we have 
$x_1=8$, $y_1=-1$, $x_2=4$, $y_2=-1$, $x_3=16$ and $y_3=-3$.
The criterion for thermal instability of optically thick disks can
be written as:
\beq
\delta_{\Sigma T}>6-\frac{2}{q}.
\eeq

If the otically thick disk is radiative cooling dominated, above
inequality
becomes $\delta_{\Sigma T}>-\infty$, which is satisfied without any
doubt. Therefore, this indicates that the optically thick,
radiation pressure and radiative cooling dominated disk is thermally 
unstable, which was first proved by a linear study of Shakura \& 
Sunyaev $^3$. If the optically thick disk is advection-dominated,
the thermal instability criterion requires $\delta_{\Sigma T}>4$.
A recent study of Wu $^{14}$ pointed out that for an optically thick,
advection-dominated disk we have 
$\delta_{\Sigma T}\sim 8$ in the short wavelength perturbation case, 
which is the result of the fact that there
is no appreciable vertical integrated pressure change associated
with the temperature perturbations.
Therefore we still obtain that the optically thick,
advection-dominated
disk is thermally unstable, even if the effects of thermal diffusion 
and magnetic
fields are not included.

In addition, we note that in the long wavelength perturbation limit where
we have $\delta_{\Sigma T} \sim 0$, the optically thick disk may become
thermally stable when $q > \frac{1}{3}$. This also agrees well with the
implication of the turning point from the radiative cooling dominated
branch to the advection-dominated branch in the thermal equilibrium curve $^{10}$.

 The generalized criterion
can be applied to investigate the thermal stability properties of both
radiative cooling dominated disks and advection-dominated disks. The
results qualitatively agree well to those obtained by some recent detailed stability
analyses, which shows that an optically thin advection-dominated disk
is
thermally stable while an optically thick advection-dominated disk is
thermally unstable against the short wavelength perturbations $^{12,13}$.
 Moreover, without solving the detailed disk structure, our new criterion can be 
used to check whether ot not the
disk is thermally stable when the advection effect and the different cooling mechanism are involved.

We note that our new  criterion depends on the relative surface density
change $\delta_{\Sigma T}$, which is a function of the perturbation wavelength
 and other disk parameters such as viscosity, Mach number, disk height and
advection term $^{14}$. In the limit of short or long wavelength perturbations, we can 
easily obtain the thermal stability properties of the 
disks from the criterion by adopting the known values of $\delta_{\Sigma T}$.
However, in the case of intermediate perturbation wavelength, we have to obtain
$\delta_{\Sigma T}$ first by numerically solving the perturbed equations before we adopt 
the new criterion. In this sense, we think that the generalized thermal instability
criterion suggested in this letter is only appropriate in the short and long perturbation
wavelength limits and the quantitative stability
analyses of accretion disks are still needed for the case of intermediate perturbation wavelength.

I thank Dr. Xingming Chen at Lick Observatory for helpful
discussions.\\[7mm]
{\bf REFERENCES}
\begin{description}
\itemsep -0.8mm
\item 1. J.E. Pringle, Ann. Rev. Astro. Astrophy., 19(1981), 137
\item 2. N.I. Shakura, R.A. Sunyaev, Astro. Astrophy., 24(1973), 337
\item 3. N.I. Shakura, R.A. Sunyaev, Mon. Not. Royal. Astro. Soc., 175(1976), 613
\item 4. A.P. Lightman, D.N. Eardley, Astrophy. J., 187(1974), L1
\item 5. S.L. Shapiro, A.P. Lightman, D.N Eardley, Astrophy. J., 204(1976), 187
\item 6. J.E. Pringle, Mon. Not. Royal. Astro. Soc., 177(1976), 65
\item 7. T. Piran, Astrophy. J., 221(1978), 652
\item 8. J. Frank, A.R. King, D.J. Raine, {\it Accretion Power in Astrophysics
(2nd edition)}, Cambridge University Press (1992)
\item 9. R. Narayan, I.Yi, Astrophy. J., 428(1994), L13
\item 10. M.A. Abramowicz, \etal, Astrophy. J., 438(1995), L37
\item 11. X. Chen, \etal, Astrophy. J., 443(1995), L61
\item 12. S. Kato, M.A. Abramowicz, X. Chen, Pub. Astro. Soc. Japan, 48(1996), 67
\item 13. X.B. Wu, Q.B. Li, Astrophy. J., 469(1996), 776
\item 14. X.B. Wu, Astrophy. J., 489(1997), 222
\item 15. X.B. Wu, Mon. Not. Royal. Astro. Soc.,  292(1997), 113
\end{description}

\end{document}